\documentclass{PoS}
\newcommand \hmu {\hat{\mu}}

\title{The strange degrees of freedom in QCD at high temperature}

\ShortTitle{The strange degrees of freedom in QCD at high temperature}

\author{\speaker{Christian Schmidt} (for RBC-Bielefeld
  Collaboration)\thanks{CS acknowledges support by the BMBF under
    grant 05P12PBCTA, the numerical calculations have been performed
    using the USQCD GPU-clusters at JLab, the Bielefeld GPU cluster
    and the NYBlue at the NYCCS} \\ Universit\"at Bielefeld,
  Fakult\"at f\"ur Physik, Postfach 100131, D-33501 Bielefeld,
  Germany\\ E-mail: \email{schmidt@physik.uni-bielefeld.de}}

\abstract{We discuss recent results on fluctuations of conserved
  charges, and their approach to the hadron resonance gas at
  temperatures below the chiral crossover ($T_c$) as well as to a gas
  of free quarks at very high temperatures.  We will focus on the
  strange degrees of freedom and verify that they are consistent with
  those of an uncorrelated gas of hadrons for temperatures of $T\lesssim 160$~MeV and that they can
  be described by a quasi particle model at best for $T\gtrsim 2T_c$.  
  To extract this information we use all cumulants of net
  baryon number and net strangeness fluctuations and their correlations 
  up to the fourth
  order. In particular we propose observables that serve as indicator
  for the validity of the hadronic degrees of freedom and show that
  the partial pressures from different strangeness sectors agree
  separately with HRG model predicions.  }

\FullConference{31st International Symposium on Lattice Field Theory - LATTICE 2013\\
		July 29 - August 3, 2013\\
		Mainz, Germany}

\begin{document}

\section{Introduction}
The fluctuations of conserved charges in hot and dense QCD carry a
variety of information that can be used to probe thermodynamic 
properties of QCD. These quantities can be obtained as derivatives 
of the QCD partition function with respect to appropriate chemical 
potentials. The original motivation to study these quantities stems
from the idea of calculating successive orders in a Taylor expansion
of the pressure with respect to a baryon chemical potential \cite{BS1}.
As direct lattice QCD calculations are harmed by the notorious sign
problem, a Taylor expansion provides one possibility to obtain corrections
to bulk thermodynamic quantities that are due to the presence of a dense 
medium, in a rigorous and controled manner. Moreover, the convergence radius of such 
expansions can provide valuable information on the existence or non-existence
of a critical point in the QCD phase diagram \cite{CS,GG}. Secondly, 
cumulants of fluctuations of conserved charges can also be measured experimentally in 
ultra relativistic heavy ion collisions (HICs).
They are excellent indicators for QCD critical behavior \cite{Stephanov} and 
thus important observables in the context of a critical point search.
In fact, recently large experimental effort is made, {\it e.g.} as part of the 
RHIC beam energy scan (BES) \cite{STAR, PHENIX}, to determine higher order
cumulants of multiplicity fluctuations. During the evolution of a HIC, a point 
is reached where particle multiplicities and their fluctuations freeze-out, once the 
fireball that is created by the colliding nuclei, has sufficiently cooled and expanded.
Matching (lattice) QCD calculations and experimental measurements on fluctuations, one 
is able to estimate external thermodynamic parameters, such as temperature
and chemical potential, that are associated with the freeze-out point. \cite{freeze}. 

Finally --- and this is the motivation
that builds the basis of this study --- cumulants of conserved charge fluctuations
reveal information of the elementary units of charges that are carried by the 
effective degrees of freedom in the system \cite{EKR}. With other words, the assumption 
that the quantum numbers of the effective degrees of freedom in the system are those of hadrons
will lead to very different predictions for the fluctuations than 
those one would get from a gas of free quarks. For our lattice calculations we
use $2+1$ flavor of highly improved staggered quarks (HISQ) \cite{HISQ} 
with almost physical 
masses, {\it i.e.} the light quark mass ($m_l$) has been chosen as $m_l=m_s/20$, 
while we tuned the strange quark mass ($m_s$) to its physical value. This corresponds to 
a Goldstone Pion mass of $m_\pi\approx 160$~MeV. We will compare our lattice QCD
results on fluctuations to the hadron resonance gas (HRG) model 
at low temperatures and to quasi free quarks at high temperatures. Here we focus 
on net strangeness fluctuations and their correlations with net baryon number. 
As its mass is of the order of the temperatures of interest, the strange quark always played a special role in the 
analysis of the quark gluon plasma, in both experimental and theoretical 
studies \cite{hic-rev}. Furthermore, the possibility of a flavor specific 
hierarchy of deconfinement has been discussed recently \cite{Ratti}, triggered
by rather large values of the transition temperature, determined from Polyakov Loop and 
net strangeness fluctuations \cite{WB-Tc}. We do not find a clear signal for such 
hierarchy in this study \cite{our}.

\section{Definition and results of cumulants}
\begin{figure}[t]
\begin{center}
\begin{minipage}{8.8pc}
\includegraphics[width=8.8pc]{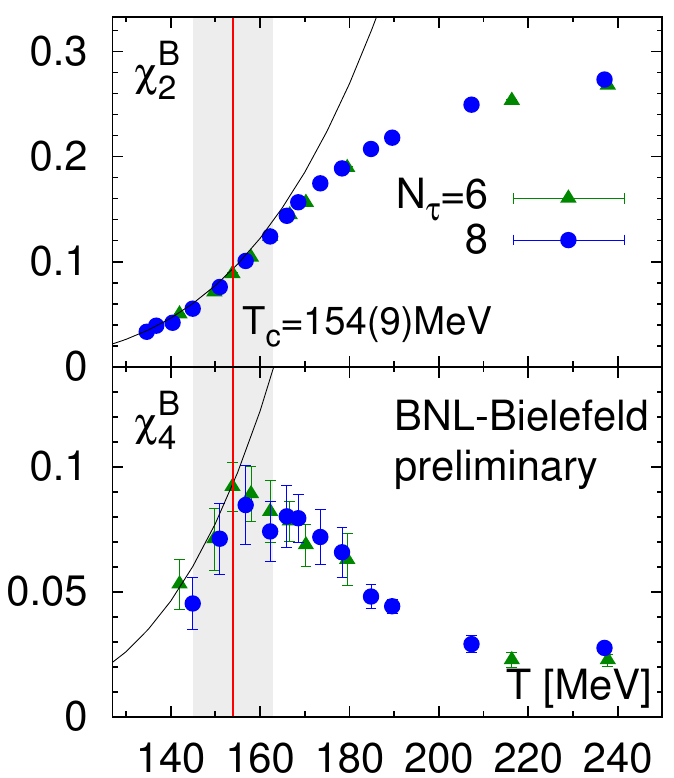}
\end{minipage}
\begin{minipage}{8.8pc}
\includegraphics[width=8.8pc]{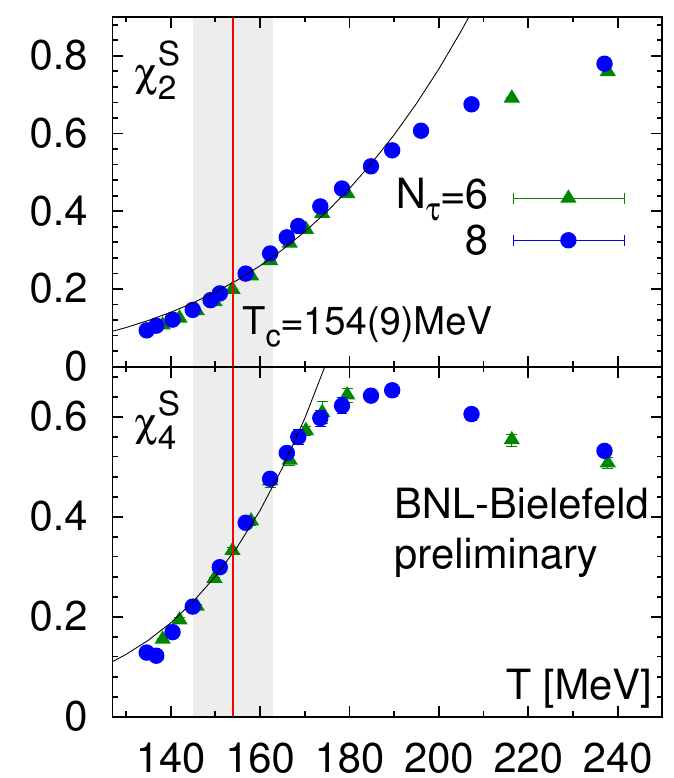}
\end{minipage} 
\begin{minipage}{8.8pc}
\includegraphics[width=8.8pc]{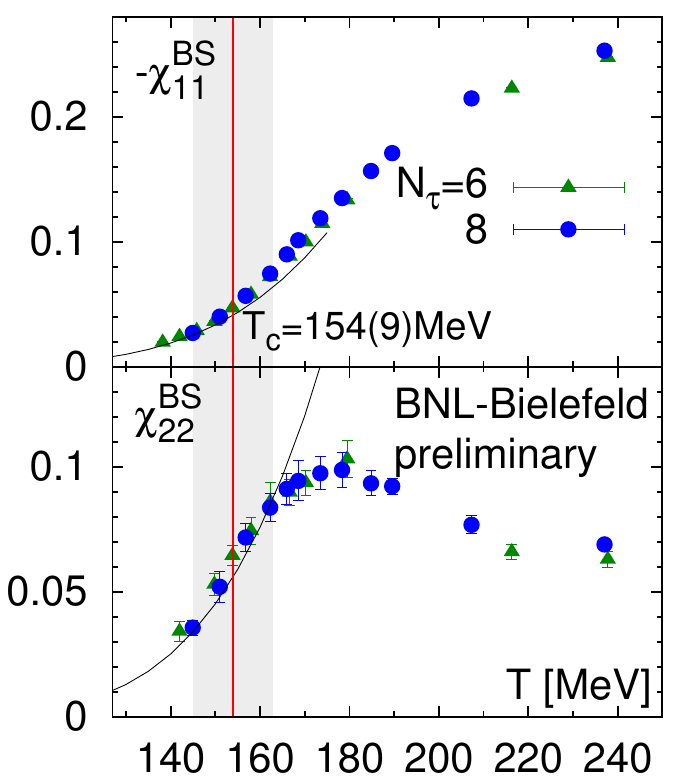}
\end{minipage}
\begin{minipage}{8.8pc}
\includegraphics[width=8.8pc]{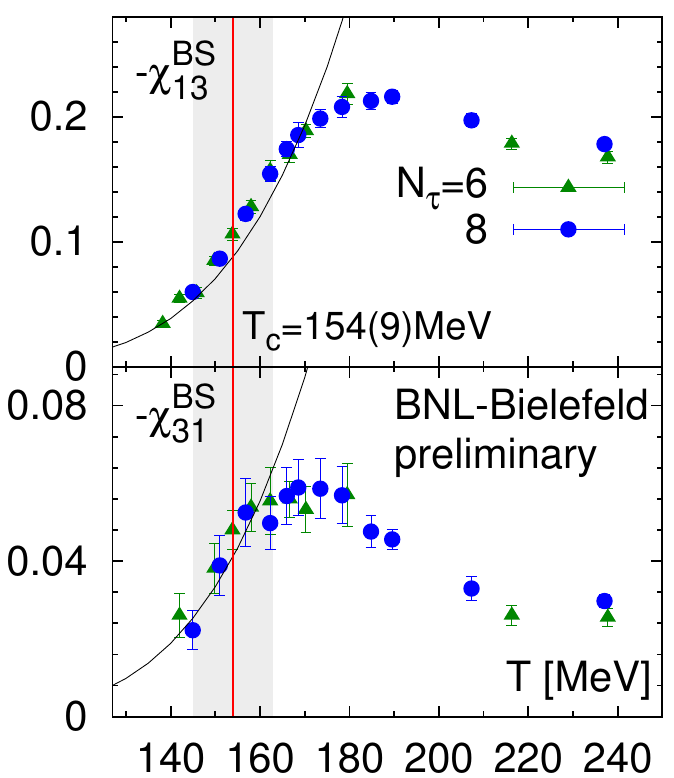}
\end{minipage} 
\end{center}
\caption{Diagonal cumulants of net baryon number and net strangeness
  fluctuations and their off-diagonal correlations up to 4th order
  as a function of temperature. Vertical bands indicate the chiral
  crossover temperature \cite{hotQCDTc}. The corresponding results
  from the HRG model are shown as solid black lines.\label{fig:fluct}}
\end{figure}

In thermodynamics, the study of susceptibilities,
i.e. derivatives of the logarithm of the partition function with
respect to external parameter, is a widely used concept, as the
calculation of the partition function itself is in many cases rather
cumbersome. In the framework of lattice regularized QCD, cumulants of
the fluctuations of conserved charges, such as net baryon number (B),
net strangeness (S) or net electric charge (Q), can be readily
obtained as such (generalized) susceptibilities as they are defined as
derivatives with respect to the chemical potentials of the conserved
charges $\mu_X$ with $X=B,Q,S$. The diagonal cumulants and their
off-diagonal correlations can be defined as
\begin{equation}
\chi_{mnl}^{BQS} = \left. \frac{\partial^{(m+n+l)} [P(T,\hmu_B,\hmu_Q,\hmu_S)/T^4]} 
{\partial \hmu_B^m \partial \hmu_Q^n \partial \hmu_S^l }\right|_{\vec{\mu}=0}\;,
\label{eq:chi}
\end{equation}   
where $P$ denotes the pressure\footnote{In the grand canonical
  ensemble we have $P=T\ln Z/V$} and $\hmu_X=\mu_X/T$. In the
following we will drop pairs of upper and lower indices in case the
lower index vanish identical, {\it i.e.} we follow the convention
$\chi_{101}^{BQS}\equiv\chi_{11}^{BS}$ and so forth. Experimentally, the
cumulants are determined from corresponding net charge %($N_X$)
fluctuations around their central values. %{\it E.g.}
%diagonal cumulants up to the forth order are given by  
%\begin{eqnarray}
%\left(VT^3\right)\cdot\chi^X_2 &=& \left<\left(\delta N_X\right)^2\right>\;,\\
%\left(VT^3\right)\cdot\chi^X_4 &=& \left<\left(\delta N_X\right)^4\right>-3\left<\left(\delta N_X\right)^2\right>^2\;,
%\end{eqnarray}
%with volume $V$, temperature $T$ and $\delta N_X=N_X-<N_X>$.
In Fig.~\ref{fig:fluct} we show our lattice results for all net baryon
number and net strangeness cumulants and their correlations up to
fourth order, obtained from $\mathcal{O}(10^4-10^5)$ lattice
configurations per temperature.
A more detailed description of the necessary operators that have to be
measures can be found, {\it e.g.}, in \cite{six}. Also shown are the
corresponding results from the HRG model, that will be discussed in
more detail in Sec.~\ref{sec:lowT}. In general we find good agreement
with the HRG model for temperatures $T\lesssim 160$ MeV.

\section{The hadronic phase \label{sec:lowT}}
\begin{figure}[t]
\begin{center}
\begin{minipage}{14pc}
\includegraphics[width=14pc]{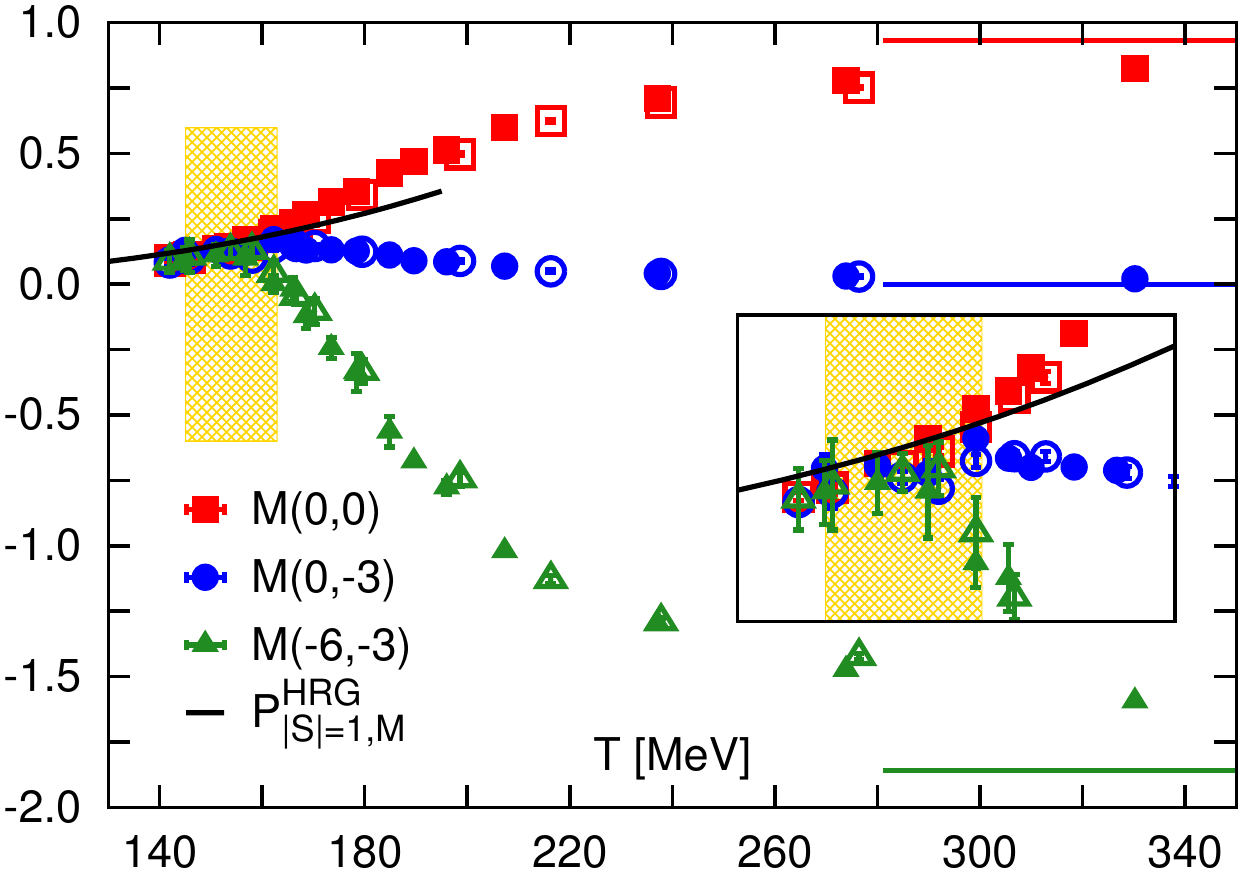}
\end{minipage}\hspace{1pc}%
\begin{minipage}{14pc}
\includegraphics[width=14pc]{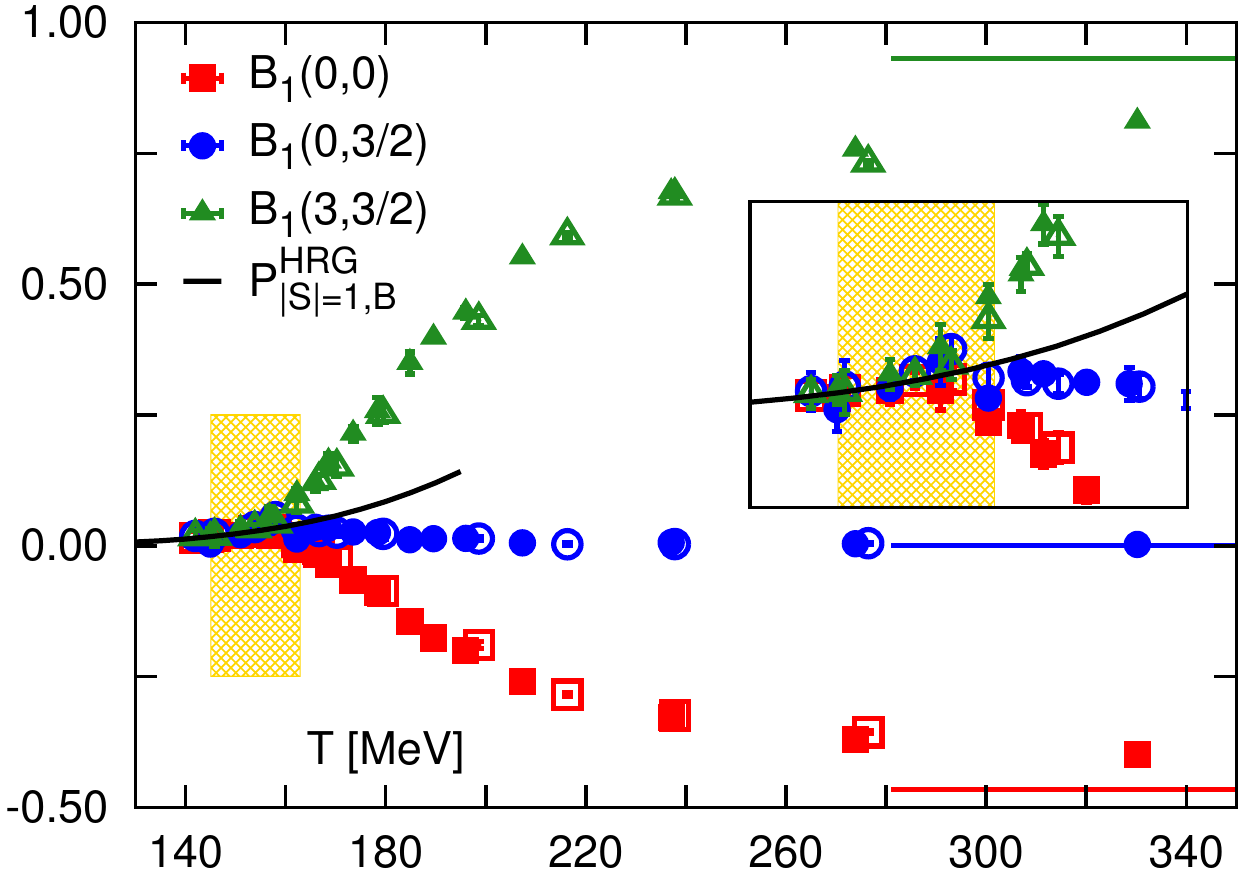}
\end{minipage} 
\begin{minipage}{14pc}
\includegraphics[width=14pc]{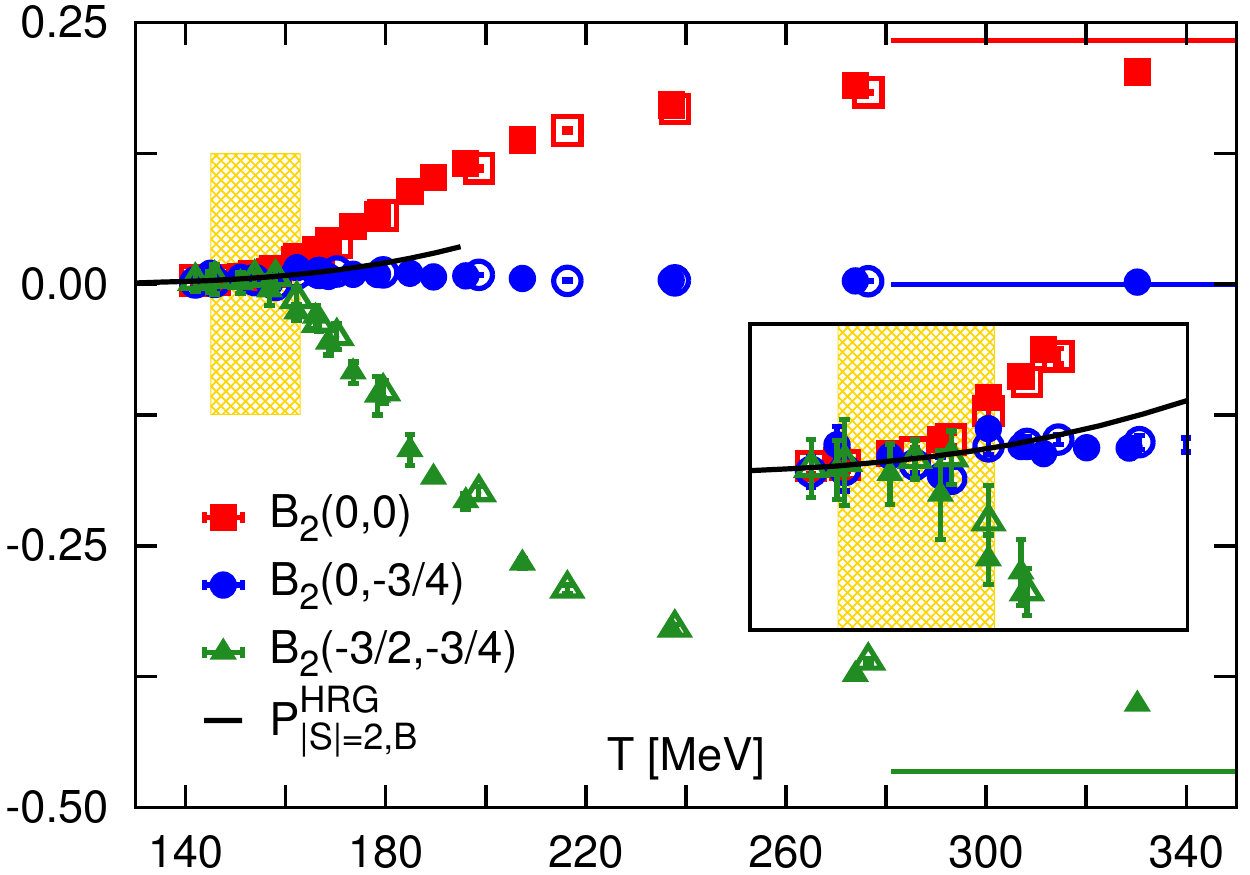}
\end{minipage}\hspace{1pc}%
\begin{minipage}{14pc}
\includegraphics[width=14pc]{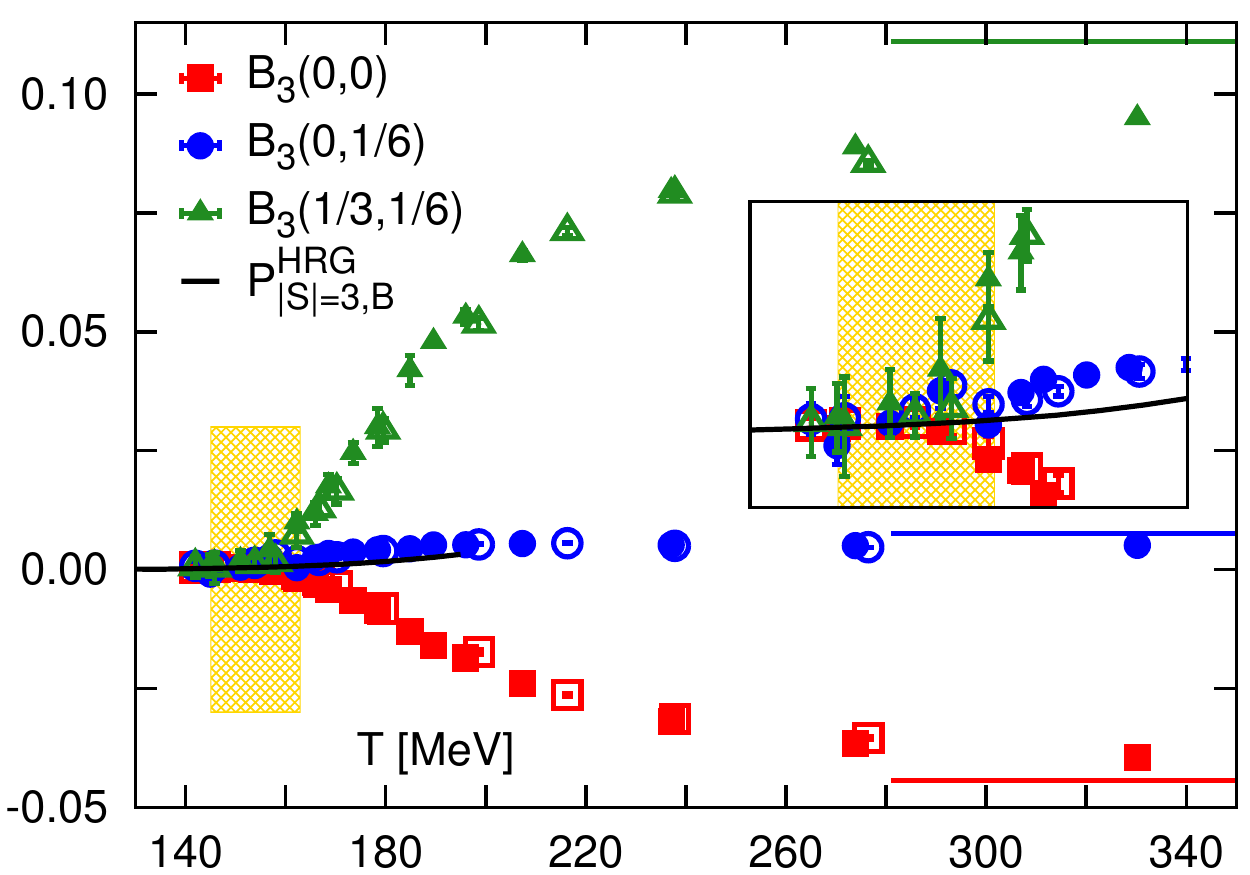}
\end{minipage} 
\end{center}
\caption{Operators that project for low temperatures onto the partial
  pressure of strange mesons (top-left), baryons with strangeness
  $|S|=1$ (top-right), $|S|=2$ (bottom-left) and $|S|=3$
  (bottom-right). Results from $N_\tau=6$ and $8$ lattices are shown
  by open and full symbols, respectively.  Also shown are results from
  the HRG model (black solid lines). Vertical bands indicate the
  chiral crossover temperature \cite{hotQCDTc}. \label{fig:pp}}
\end{figure}

\begin{figure}[t]
\begin{center}
\includegraphics[width=30pc]{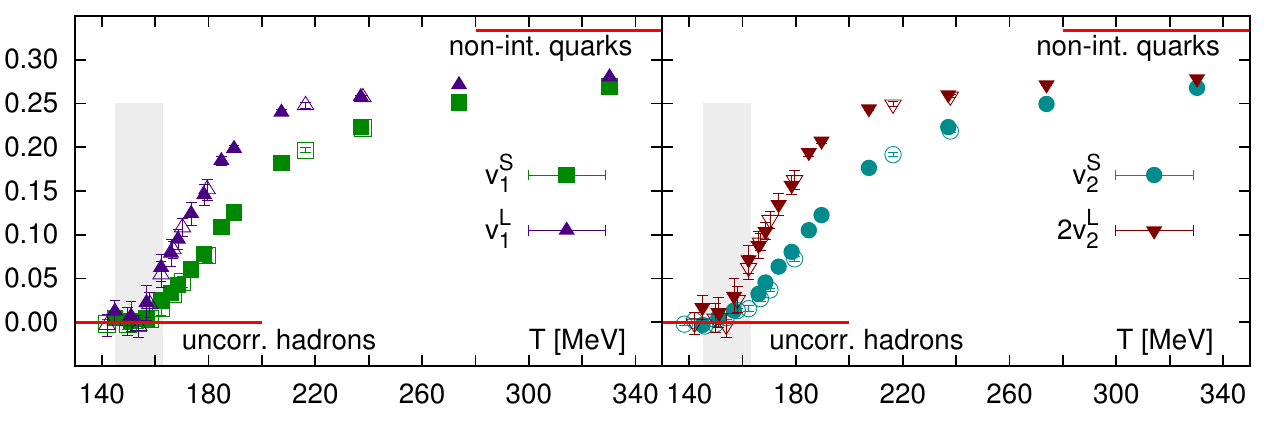}
\end{center}
\caption{Quantities that vanish within and uncorrelated gas of hadrons (Boltzmann approximation). \label{fig:v}}
\end{figure}

A widely used model for the hadronic phase of QCD is the HRG
model. Here we use the most simplistic version where the total pressure
is given as sum over all partial pressures that are obtained by the
free quantum gases of all particles and resonances from the particle
data booklet, up to a mass cut-off of $2.5$~GeV. {\it I.e.}, no
interactions (for instance hard core repulsion) are taken into
account. In fact, the interactions between hadrons
are already included through the appearance of resonances
\cite{DMB}. Such type of models are used very successfully for the
description of experimentally measured particle multiplicities and
their fluctuations \cite{HRG} and also yield important guiding for bulk
thermodynamic observables from the lattice
\cite{HRG-eos,hotQCDHRG}. From Fig.~\ref{fig:fluct} it is already clear
that the HRG model leads to an excellent description also for the
observables presented here. This statement can, however, be further
refined. Firstly, we can check whether the agreement with the HRG is
valid within separate strangeness sectors. Secondly, we can construct
observables that vanish independent of temperature (up to corrections
to Boltzmann statistics) as soon as the HRG yields a valid description
of the system. Those quantities provide an optimal measure for the
validity of the HRG model. Within the HRG model the pressure that stem
from strange particles receives contributions from 4 different
sectors,
\begin{eqnarray}
P^{\rm HRG}_S(T,\hmu_B,\hmu_S) &=& P^{\rm HRG}_{|S|=1,M}(T) \cosh(\hmu_S) 
+ P^{HRG}_{|S|=1,B}(T) \cosh(\hmu_B-\hmu_S)
\nonumber \\
&+& P^{HRG}_{|S|=2,B}(T) \cosh(\hmu_B-2\hmu_S)
+ P^{HRG}_{|S|=3,B}(T) \cosh(\hmu_B-3\hmu_S) \;,
\label{eq:p}
\end{eqnarray}
where $P^{HRG}_{|S|=i,M/B}$ denotes the partial pressure from mesons
($M$) or baryons ($B$) with strangeness $|S|=i$. We also assume
Boltzmann statistics here, which leads to the factorization of the
$\mu$ dependence\footnote{The validity of Boltzmann approximation is
  ensured here, since for the lightest strange hadron (Kaon) we have
  $m/T\gtrsim 3$ for the temperature values considered
  here.}. Eq.~(\ref{eq:p}) immediately leads to linear relations between
the quantities defined through Eq.~(\ref{eq:p}) (shown in
Fig.~\ref{fig:fluct}) and the partial pressures from different
strangeness sectors. Inverting these relations enables us to project to
the partial pressures. we denote the projectors as $M_1,B_1,B_2$ and
$B_3$, which we plot in Fig.~\ref{fig:pp}
Note that the projectors are not defined uniquely, as each of it
depends on two parameter $c_1,c_2$. However, as long as the HRG is
valid, the dependence on $c_1$ and $c_2$ should vanish. This is seen
in Fig.~\ref{fig:pp} where we plot each of the projectors for 3
choices of the parameters. We find indeed, that for $T\lesssim
160$~MeV the dependence on $c_1$ and $c_2$ vanish and the results agree
with each other and also with the partial pressure of the specific
strange sector that is computed from the HRG model. The 2-dim. freedom
in the definition of the projectors originates from the 2-dim. null
space of the linear mapping between our 6 basis observables shown in
Fig.~\ref{fig:fluct} and the partial pressures from the 4 strangeness
sectors. As basis vectors of this null space we can choose the linear combinations $v_1^S =
\chi_{31}^{BS} - \chi_{11}^{BS}$ and $v_2^S = \frac{1}{3} (\chi_2^S -
\chi_4^S ) - 2 \chi_{13}^{BS} - 4 \chi_{22}^{BS} - 2 \chi_{31}^{BS}$ 
\cite{our}. Again, these combinations of cumulants vanish in 
the HRG model (up to corrections to the Boltzmann statistics) and provide
a measure for the quality of the HRG model. As $v_1^S$ and $v_2^S$ receive 
contributions from strange hadrons only, they also indicate the temperature from where on 
strange quarks start to be liberated into the system. The whole 
analysis can be repeated for correlations of the net baryon number with 
other flavors as proposed in \cite{Ratti}. In Fig.~\ref{fig:v} we plot the
susceptibilities $v_1^S, v_2^S$ together with $v_1^L, v_2^L$, where $L$
indicates the net lightness number. 
In analogy to Eq.~(\ref{eq:chi}) we
define $\chi_n^L$ by $\chi_n^L=\partial^n (p/T^4)/\partial \hmu_L^n$ and 
$\partial \hmu_L=(\partial \hmu_u + \partial \hmu_d)/2$ where $\mu_u$ and $\mu_d$ 
are up and down quark chemical potentials, respectively. 
In contrast to $v_1^S$ and $v_2^S$,
$v_1^L$ and $v_2^L$ receive also contributions from protons and neutrons and thus can be understood
as a more general deconfinement observable. As can be seen from Fig.~\ref{fig:v} their behavior
is very similar. Especially the temperature at which the observables deviate from zero is similar. 
We thus conclude that within our current accuracy strange hadrons do not dissolve at larger temperatures
than light hadrons.

\section{The plasma phase}
\begin{figure}[t]
\begin{center}
\includegraphics[width=30pc]{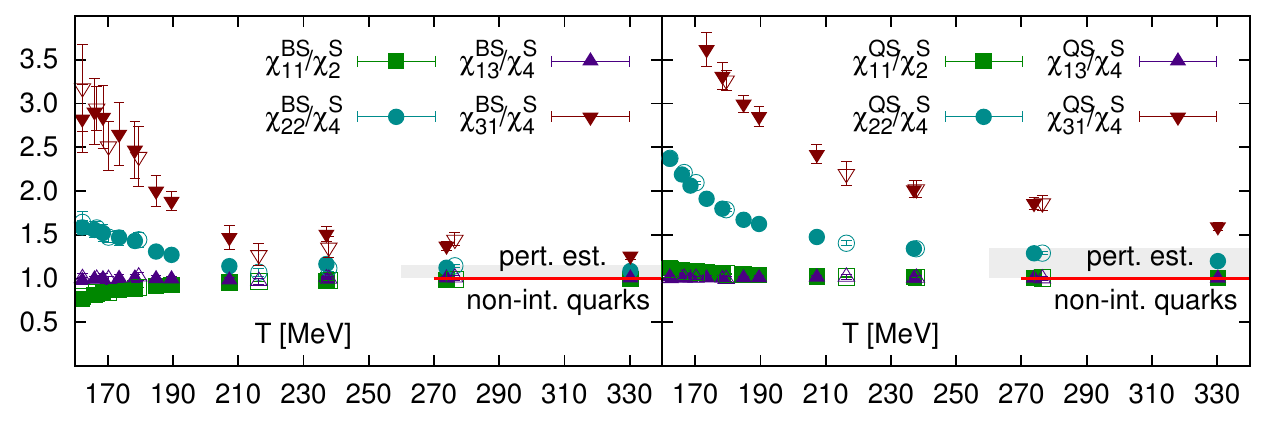}
\end{center}
\caption{Baryon number strangeness (left) and electric charge strangeness correlations (right),
properly scaled by the diagonal strangeness cumulants and powers of the fractional baryonic
and electric charges [see Eq. (4.1)].
In the non-interacting massless quark gas all these observables are unity (shown by the lines at high temperatures).
The shaded regions indicate the range of perturbative estimates (see text) for all these observables obtained using
one-loop re-summed HTL calculations \cite{HTL}. Results from $N_\tau=6$ and $8$ lattices are shown by open and filled
symbols, respectively.\label{fig:highT}}
\end{figure}

At very high temperatures we expect to find quantum numbers of the effective degrees of freedom 
that resemble those of quarks. If the strange degrees of freedom can be described by a weakly interacting 
gas of quasi-quarks, then strangeness is associated with a fractional baryon number of $B=1/3$ and a 
electric charge of $Q=-1/3$. For correlations of net strangeness with net baryon number and net electric
charge on thus obtains
\begin{equation}
\frac{\chi_{mn}^{BS}}{\chi_{m+n}^S}=\frac{(-1)^n}{3^m}\;\qquad
\frac{\chi_{mn}^{QS}}{\chi_{m+n}^S}=\frac{(-1)^{m+n}}{3^m}\;,
\end{equation}
where $n,m>0$ and $m+n=2,4$. In Fig.~\ref{fig:highT} we show our results for these ratios, scaled by the 
propper powers of fractional baryonic and electric charges.
The shaded regions at high temperatures indicate the range of values
for these ratios as predicted for weakly interacting quasi-quarks
from the re-summed Hard Thermal Loop (HTL) perturbation theory at the 
one-loop order \cite{HTL}, using one-loop running coupling obtained at
the scale between $\pi T$ and $4\pi T$. We find that second order ratios
are much closer to those expected for weakly interacting quasi-quarks, differing at a few percent level for $T\sim 1.25 T_c$. Previous
lattice QCD calculations showed similar results \cite{hotQCDHRG,WB-fluct,p4},
suggesting that the strange degrees of freedom in the QGP can be described
by weakly interacting quasi-quarks even down to temperatures very close to 
$T_c$. However, our results involving correlations of strangeness with higher
powers of baryon number and electric charge clearly indicate that a description
in terms of weakly interacting quasi-quarks can only be valid for temperatures
$T\gtrsim 2T_c$. Note, that the HTL perturbative expansion for ratios 
involving one derivative w.r.t the baryonic/electric chemical potential 
($\chi_{11}^{XS}/\chi_2^S$ and $\chi_{13}^{XS}/\chi_4^S$, $X=B,Q$)
starts differing from the non-interacting quark gas limit at 
$\mathcal{O}(\alpha_s^3\ln \alpha_s)$ \cite{pt}. For ratios involving
higher derivatives w.r.t the baryonic/electric chemical potentials
($\chi_{22}^{XS}/\chi_4^S$ and $\chi_{31}^{XS}/\chi_4^S$, $X=B,Q$)
the difference starts at $\mathcal{O}(\alpha_s^{3/2})$ \cite{HTL}.
Here $\alpha_s$ denotes the strong coupling constant. The enhancement
of the higher order baryon number strangeness and electric charge 
strangeness correlations is thus likely to be expected within the 
regime of the validity of the weak coupling expansion.


\begin{thebibliography}{99}
\bibitem{BS1}
  C.~R.~Allton, S.~Ejiri, S.~J.~Hands, O.~Kaczmarek, F.~Karsch, E.~Laermann, C.~Schmidt and L.~Scorzato,
  %``The QCD thermal phase transition in the presence of a small chemical potential,''
  Phys.\ Rev.\ D {\bf 66} (2002) 074507
  [hep-lat/0204010].
\bibitem{CS}
  C.~Schmidt,
  %``Net-baryon number fluctuations in (2+1)-flavor QCD,''
  Prog.\ Theor.\ Phys.\ Suppl.\  {\bf 186} (2010) 563
  [arXiv:1007.5164 [hep-lat]].
\bibitem{GG}
  R.~V.~Gavai and S.~Gupta,
  %``The Critical end point of QCD,''
  Phys.\ Rev.\ D {\bf 71} (2005) 114014
  [hep-lat/0412035].
\bibitem{Stephanov}
  M.~A.~Stephanov,
  %``Non-Gaussian fluctuations near the QCD critical point,''
  Phys.\ Rev.\ Lett.\  {\bf 102} (2009) 032301
  [arXiv:0809.3450 [hep-ph]].
\bibitem{STAR}
 L.~Adamczyk {\it et al.}  [STAR Collaboration],
  %``Energy Dependence of Moments of Net-proton Multiplicity Distributions at RHIC,''
  arXiv:1309.5681 [nucl-ex].
\bibitem{PHENIX}
  P.~Garg [for PHENIX Collaboration],
  %``Moments of net-charge multiplicity distribution in Au+Au collisions measured by the PHENIX experiment at RHIC,''
  PoS CPOD {\bf 2013} (2013) 050
  [arXiv:1305.7327 [nucl-ex]].  arXiv:1309.5681 [nucl-ex].
\bibitem{freeze}
  A.~Bazavov, H.~T.~Ding, P.~Hegde, O.~Kaczmarek, F.~Karsch, E.~Laermann, S.~Mukherjee and P.~Petreczky {\it et al.},
  %``Freeze-out Conditions in Heavy Ion Collisions from QCD Thermodynamics,''
  Phys.\ Rev.\ Lett.\  {\bf 109} (2012) 192302
  [arXiv:1208.1220 [hep-lat]].
\bibitem{EKR}
  S.~Ejiri, F.~Karsch and K.~Redlich,
  %``Hadronic fluctuations at the QCD phase transition,''
  Phys.\ Lett.\ B {\bf 633} (2006) 275
  [hep-ph/0509051].
\bibitem{HISQ}
  E.~Follana {\it et al.}  [HPQCD and UKQCD Collaborations],
  %``Highly improved staggered quarks on the lattice, with applications to charm physics,''
  Phys.\ Rev.\ D {\bf 75} (2007) 054502
  [hep-lat/0610092].
\bibitem{hic-rev} 
For recent reviews see: 
B. V. Jacak and B. Muller, Science {\bf 337}, 310 (2012); 
B. Jacak and P. Steinberg, Phys.\ Today {\bf 63N5}, 39 (2010).
\bibitem{Ratti} 
C. Ratti, R. Bellwied, M. Cristoforetti and M. Barbaro, Phys.\ Rev.\ D {\bf 85},
014004 (2012);  R.~Bellwied, S.~Borsanyi, Z.~Fodor, S.~DKatz and C.~Ratti, 
arXiv:1305.6297 [hep-lat].
\bibitem{WB-Tc}
Y. Aoki, Z. Fodor, S. D. Katz and K. K. Szabo, Phys.\ Lett.\ B {\bf 643}, 46 (2006);
Y. Aoki {\it et al.}, JHEP {\bf 0906}, 088 (2009);
S. Borsanyi {\it et al.}  [Wuppertal-Budapest Collaboration], JHEP {\bf 1009}, 073
(2010);
\bibitem{our} 
 A.~Bazavov, H.~-T.~Ding, P.~Hegde, O.~Kaczmarek, F.~Karsch, E.~Laermann, Y.~Maezawa and S.~Mukherjee {\it et al.},
  %``Strangeness at high temperatures: from hadrons to quarks,''
  Phys.\  Rev.\  Lett.\  {\bf 111}, 082301 (2013)
  [arXiv:1304.7220 [hep-lat]].
\bibitem{hotQCDTc} A.~Bazavov, T.~Bhattacharya, M.~Cheng, C.~DeTar, H.~T.~Ding, S.~Gottlieb, R.~Gupta and P.~Hegde {\it et al.},
  %``The chiral and deconfinement aspects of the QCD transition,''
  Phys.\ Rev.\ D {\bf 85} (2012) 054503
  [arXiv:1111.1710 [hep-lat]].
\bibitem{six}
  C.~R.~Allton, M.~Doring, S.~Ejiri, S.~J.~Hands, O.~Kaczmarek, F.~Karsch, E.~Laermann and K.~Redlich,
  %``Thermodynamics of two flavor QCD to sixth order in quark chemical potential,''
  Phys.\ Rev.\ D {\bf 71} (2005) 054508
  [hep-lat/0501030].
\bibitem{DMB}
  R.~Dashen, S.~-K.~Ma and H.~J.~Bernstein,
  %``S Matrix formulation of statistical mechanics,''
  Phys.\ Rev.\  {\bf 187} (1969) 345.
\bibitem{HRG}
  P.~Braun-Munzinger, K.~Redlich and J.~Stachel,
  %``Particle production in heavy ion collisions,''
  In *Hwa, R.C. (ed.) et al.: Quark gluon plasma* 491-599
  [nucl-th/0304013].
\bibitem{HRG-eos}
  M.~Cheng, S.~Ejiri, P.~Hegde, F.~Karsch, O.~Kaczmarek, E.~Laermann, R.~D.~Mawhinney and C.~Miao {\it et al.},
  %``Equation of State for physical quark masses,''
  Phys.\ Rev.\ D {\bf 81} (2010) 054504
  [arXiv:0911.2215 [hep-lat]].
\bibitem{hotQCDHRG}
 A.~Bazavov {\it et al.}  [HotQCD Collaboration],
  %``Fluctuations and Correlations of net baryon number, electric charge, and strangeness: A comparison of lattice QCD results with the hadron resonance gas model,''
  Phys.\ Rev.\ D {\bf 86} (2012) 034509
  [arXiv:1203.0784 [hep-lat]].
\bibitem{HTL}
J.~O.~Andersen, S.~Mogliacci, N.~Su and A.~Vuorinen,
  %``Quark number susceptibilities from resummed perturbation theory,''
  Phys.\ Rev.\ D {\bf 87} (2013) 074003
  [arXiv:1210.0912 [hep-ph]].
\bibitem{WB-fluct} 
 S.~Borsanyi, Z.~Fodor, S.~D.~Katz, S.~Krieg, C.~Ratti and K.~Szabo,
  %``Fluctuations of conserved charges at finite temperature from lattice QCD,''
  JHEP {\bf 1201} (2012) 138
  [arXiv:1112.4416 [hep-lat]].
\bibitem{p4}
M.~Cheng, P.~Hendge, C.~Jung, F.~Karsch, O.~Kaczmarek, E.~Laermann, R.~D.~Mawhinney and C.~Miao {\it et al.},
  %``Baryon Number, Strangeness and Electric Charge Fluctuations in QCD at High Temperature,''
  Phys.\ Rev.\ D {\bf 79} (2009) 074505
  [arXiv:0811.1006 [hep-lat]].
\bibitem{pt}
 J.-P. Blaizot, E.~Iancu and A.~Rebhan, 
 Phys.\ Lett.\ B {\bf 523} (2001) 143.
\end{thebibliography}
\end{document}